\documentclass{elsart5p}

\usepackage[usenames]{color}
\usepackage{graphicx}
\usepackage{ifpdf}
\usepackage{amsmath}
\usepackage{amssymb}
\usepackage{epsfig}
\usepackage{natbib}
\newcommand{\aaa}{\hspace{0.25cm}}

\def\pbar{$\overline{p}$}
\def\error{\textit{\textsf{error}}}
\def\dbar{$\overline{d}$}

\begin{document}
%
%
\begin{frontmatter}
\title{Transport of exotic anti-nuclei:I- Fast formulae for $\overline{p}$ fluxes}
\author[INFN,LPNHE]{D. Maurin\corauthref{cor}}\ead{dmaurin@lpnhe.in2p3.fr}
\author[LAPTH,univ]{R. Taillet}\ead{taillet@lapp.in2p3.fr}
\author[DIAS,LAOG]{C. Combet}\ead{ccombet@obs.ujf-grenoble.fr}
\address[INFN]{Dipartimento di Fisica Teorica, Universit\`a di Torino, Istituto Nazionale
di Fisica Nucleare, via P. Giuria 1, I--10125 Torino, Italy}
\address[LPNHE]{LPNHE, IN2P3/CNRS/Universit\'es Paris VI et Paris VII, 4 place Jussieu
Tour 33 - Rez de chauss\'ee, 75252 Paris Cedex 05, France}
\address[LAPTH]{Laboratoire de Physique Th\'eorique {\sc lapth},
             Annecy--le--Vieux, 74941, France}
\address[univ]{Universit\'e de Savoie, Chamb\'ery, 73011, France}
\address[DIAS]{Dublin Institute for Advanced Studies {\sc dias},
       5 Merrion Square, Dublin 2, Ireland}
\address[LAOG]{Laboratoire d'Astrophysique, Observatoire de Grenoble {\sc laog},
			BP 53 F-38041 Grenoble Cedex 9, France}
\corauth[cor]{Corresponding author.}
\begin{abstract}
The Galactic secondary cosmic ray anti-proton (\pbar) flux calculated with
different propagation models is fairly consistent with data, and
the associated propagation uncertainty is small. This is not the case
for any \pbar\ exotic component of the dark matter halo (see also
the companion paper;~\citealt{Maurin:2006ps}).
Detailed propagation models are mandatory if the ultimate
goal is to explain an excess. However, simpler and faster approximate formulae
for \pbar\ are an attractive alternative to quickly check that a given dark matter model
is not inconsistent with the \pbar\ observed flux. This paper provides
such formulae. 
In addition,
they could be used to put constraints on new physics in this channel,
where an extensive scan of a large parameter space
could otherwise be quite expensive in computer ressources.
\end{abstract}
\end{frontmatter}


\section{Motivation}
The first positive detection of anti-protons at the end of the
seventies \citep{1979PhRvL..43.1196G, 1981ApJ...248.1179B,1982ICRC....9..146B}
boosted the interest for indirect searches of new physics
in the low energy \pbar\ spectrum. After 25 years of efforts
and improvements both in measurements and theoretical calculations, 
the data at low energy (up to a few GeV) 
are now well accounted for \citep{2001ApJ...563..172D}.
However, at higher energy, calculations tend to predict less \pbar\ than
observed \citep{2001ApJ...561..787B} but more data are desirable to
confirm this possible trend.
 
The uncertainty of the \pbar\ standard secondary flux \citep{2001ApJ...563..172D}
mainly comes from nuclear physics,
as that of astrophysical origin is small
---less than a few percents---and is expected to
decrease even more as new B/C measurements are available.
In contrast, the \pbar\ exotic flux---from exotic sources
following the dark matter
distribution---is plagued by the degeneracy in the propagation
parameters, the corresponding uncertainty being as large as
two orders of magnitude at low energy \citep{2004PhRvD..69f3501D}.
Until propagation is better
understood and constrained, it may not be worth using 
refined---and sometimes time consuming---models
whose result will anyway be crippled by these
large uncertainties. This is especially true when one has to scan 
as efficiently as possible the huge parameter space existing, for example,
in {\sc susy} theories.

The goal of this paper is to provide approximate formulae
for the propagated \pbar\ exotic fluxes. This is done in the framework
of the diffusion/convection model with constant wind already
discussed in
\cite{2001ApJ...555..585M,2001ApJ...563..172D,2002A&A...388..676B,2004PhRvD..69f3501D}.
These formulae 
are reasonably accurate, fast to compute and easy to implement.
They have to be thought as an easy-to-use tool, for phenomenologists
interested in beyond-the-standard-model theories and wishing to quickly check that
any new physics model on the market does not violate the
\pbar\ constraint.
Once the interesting regions of the parameter space
(for the dark matter candidate) are identified,
more elaborate treatments should be implemented
(e.g. {\sc Galprop}---\citealt{2005AdSpR..35..156M};
{\sc DarkSusy}---\citealt{2004JCAP...07..008G}; or \citealt{2004PhRvD..69f3501D}).

The paper is organized as follows:
\begin{itemize}
 \item In Sec.~\ref{sec:2D}, we remind the salient ingredients
 of the constant wind/diffusion model for secondary and primary exotic anti-protons,
 using a Bessel expansion formalism. 
 \item In Sec.~\ref{sec:alternative}, two alternative formulations of the \pbar\ 
 primary exotic flux are presented.
 \item In Sec.~\ref{sec:results}, all the formulations are compared and their relative
 merits discussed.
\end{itemize}
The formulae given in Sec.~\ref{sec:alternative} are implemented in numerical
routines publicly available\footnote{http://wwwlapp.in2p3.fr/$\sim$taillet/mtc/mtc\_code.tar}.

\section{2D--model with constant wind $V_c$}
\label{sec:2D}
Cosmic ray fluxes are determined by the transport equation,
as given, e.g., in \citet{1990acr..book.....B}.
Throughout the paper, we use the so-called thin-disk approximation
where the gas is contained in a layer
of thickness $2h=200$~pc. Cylindrical symmetry is assumed,
and the diffusion coefficient is constant in the whole Galaxy,
\[
K(E)=\beta K_0{\cal R}^{\delta} \quad\quad \text{(${\cal R}=pc/|Z|e$ is the rigidity)}.
\]
A galactic wind $V_c$ of constant magnitude, directed outwards
along $z$, is also included.

\subsection{Diffusion equation}
Denoting the differential
density as $N$ leads to
\begin{eqnarray}
\label{2D-exotic-thin-disk}
 \displaystyle  \left\{ -K\triangle \right. & + &\left. V_c \frac{\partial }{\partial z} 
  + 2h \Gamma_{\rm tot}\delta(z)\right\} N \nonumber\\
  \displaystyle &+&
 2h \delta(z)\left[ b(E) \frac{dN}{dE} + c(E) \frac{d^2 N}{dE^2}\right] \nonumber \\
 &=& {\cal S}(r,z,E).
\end{eqnarray}
The quantity $\Gamma_{\rm tot}= \sum_{\rm ISM} n_{\rm ISM}.v.\sigma^{\bar{p}}_{\rm ISM}$
is the destruction rate of \pbar\ in the thin gaseous disk ($n_{\rm ISM}=$H, He).
The terms $b(E)$ and $c(E)$ correspond respectively to a drift term (coulomb,
ionization, adiabatic losses and reacceleration) and a diffusion
term (reacceleration depending on the Alfv\'enic velocity)
in energy space (e.g., \citealt{2001ApJ...555..585M,2002A&A...394.1039M}).
The quantity ${\cal S}(r,z, E)$ stands for the source term and has three contributions:
secondary (spallation-induced), tertiary (non-annihilating rescattering) and
primary (exotic dark matter)---\citet{2001ApJ...563..172D,2004PhRvD..69f3501D}.
In the remaining of the paper, we term {\em energy redistribution}
the effects of drift, diffusion and tertiary contribution.

\subsection{Propagation parameters}
The parameters of the model relevant to the present study 
are i) the size $L$ of the diffusive halo of the Galaxy, ii)
the normalization of the diffusion coefficient $K_0$, iii) its slope
$\delta$ and iv) the constant galactic wind $V_c$.
The boundary conditions are $N(z=L,r)=0$ and $N(r=R,z)=0$,
where the Galactic radius is $R=20$~kpc.
The propagation parameters have been determined
from the analysis of the B/C ratio in \citet{2001ApJ...555..585M}:
only a sub-region of the propagation
parameter space is compatible with B/C data.
The accuracy of the approximate formulae given in this work is studied for
this sub-region only. Within the latter,  the sets of parameters leading to the
maximum/minimum primary exotic fluxes (see \citealt{2004PhRvD..69f3501D}
and the companion paper) are gathered in Tab.~\ref{tab:param},
along with the configuration corresponding to
the best fit to B/C data. These three sets were previously
used in \citet{2004PhRvD..69f3501D,2005PhRvD..72f3507B}.
\begin{table}
\centering
\begin{tabular}{c c c  c c}
\hline\hline
 ~~~~Set~~~~& $\delta$ & ~$K_0$ (kpc$^2$~Myr${^{-1}}$)~ & $L$ (kpc)& ~$V_c$ (km~s$^{-1}$) \\ \hline
{\em max} & 0.46 & 0.0765 & 15 & 5\\
{\em best} & 0.7 & 0.0112 & 4 & 12 \\
{\em min} & 0.85  & 0.0016 & 1 & 13.5  \\
\hline
\end{tabular}
\caption{\label{tab:param}Propagation parameters consistent with B/C data~\citep{2001ApJ...555..585M}.
The set labeled {\em best} corresponds to the best fit to B/C data, while those
labeled {\rm min} and {\rm max}  correspond to sets which give minimum and maximum
exotic fluxes~\citep{2004PhRvD..69f3501D}.}
\end{table}

       \subsection{Secondary \pbar\ flux}
\label{app:fit}
The secondary \pbar\ flux has been studied in detail in
\citet{2001ApJ...563..172D}: every set of propagation parameters
consistent with the observed B/C ratio leads to the same
\pbar\ secondary flux, within the $\sim 10\%$ uncertainties
mostly of nuclear origin. Moreover, the \pbar\ fluxes obtained
from other models are in fair agreement (see e.g. \citealt{2002ApJ...565..280M}).

\begin{figure}[!t]
\begin{center}
\includegraphics[width=\columnwidth,angle=0,clip=]{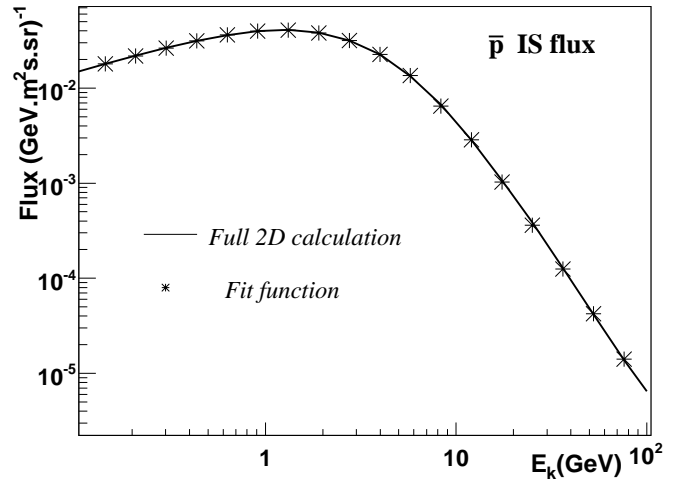}
\caption{\label{fig:pbar_fit} Reference interstellar standard \pbar\ flux along with its fit function as given
by Eq.~(\ref{eq:pbar_fit}).}
\end{center}
\end{figure}
If one wishes to exclude, for example, an exotic primary contribution
by scanning the SUSY parameter space, it is useless to calculate again
and again the same quantity with various input propagation parameters.
The following fit function, once modulated, provides a proper description of the measured
\pbar\ secondary flux,
regardless of the propagation model. Setting $x\equiv \log(E_k)$
where $E_k$ is the kinetic energy in GeV, the standard IS \pbar\ flux
in (GeV~m$^2$~s~sr)$^{-1}$ is parameterized as
\begin{eqnarray}
  \Phi^{\bar{p}~-~{\rm IS}}_{\rm standard}= \left\{
   \begin{array}{ll}
\label{eq:pbar_fit}
       \displaystyle
               \exp\left( \sum_{i=0}^{4} C_i x^i \right)
                               \quad {\rm if~} E_k<11 ~{\rm GeV}\;,\vspace{4mm}\\
\displaystyle
               \exp \left( D_0 x^{D_1} \right) \quad {\rm otherwise} \;;\vspace{2mm}
   \end{array}
   \right.
\end{eqnarray}
The constants are defined by $C_0 = -3.211$, $C_1 = 0.12145$, $C_2 = -0.2728$, $C_3 = -0.075265$, $C_4 = -0.007162$,
and $D_0=-2.02735$ and $D1 = 1.16463$. The fit is shown
in Fig.~\ref{fig:pbar_fit}.

\subsection{Exotic primary \pbar\ flux}
\label{subsec:2Dfull}
The derivation for the primary flux has been presented in~\citet{2002A&A...388..676B}\footnote{Footnote that there is a misprint in Eq.~(A.5) of \citep{2002A&A...388..676B}.
The product $KA_iS_i$ should read $KS_i$.}. The solution
can be found using a Bessel expansion (see e.g., 
\citealt{2001ApJ...555..585M})
\[
N^{\rm prim}(r,z) = \sum_{i=1}^{\infty} {\cal N}^{\rm prim}_i(z) J_0\left(\zeta_i \frac{r}{R}\right)\;,
\]
where $\zeta_i$ is the $i$-th zero of $J_0$ and
\begin{eqnarray}
{\cal N}^{\rm prim}_i(z) &=& {\cal N}_i(0) e^{\frac{V_cz}{2K}} \frac{\sinh\left[\frac{S_i}{2}(L-z)\right]}{\sinh\left(\frac{S_iL}{2}\right)}
\label{Ni_nocal}\\\nonumber
&+& e^{-\frac{V_c}{2K}(L-z)}\frac{\sinh(S_iz/2)}{\sinh(S_iL/2)} \frac{y_i(L)}{KS_i} -\frac{y_i(z)}{KS_i}.
\end{eqnarray}
Each ${\cal N}_i(0)$ obeys a differential equation
\begin{equation}
{\cal N}_i(0) = {\cal N}_i^\star (0) - \frac{2h}{K} \left[ b(E) \frac{d{\cal N}_i(0)}{dE} + c(E) \frac{d^2{\cal N}_i(0)}{dE^2}\right],
\label{eq:fuck}
\end{equation}
where ${\cal N}_i^\star (0)$ is the solution without energy redistributions:
\begin{equation}
{\cal N}_i^\star(0) \equiv e^{-V_cL/2K}\frac{y_i(L)}{A_i\sinh(S_iL/2)}\; . \label{Ni_cal}
\end{equation}
Equation~(\ref{eq:fuck}) is solved with a numerical scheme detailed in \citet{2001ApJ...563..172D}.
The tertiary contribution is not shown in the former equation although it is taken
into account in our numerical calculations.
The remaining quantities  are defined as follows:
\begin{eqnarray}
& &   y_i(z)\equiv 2\int_0^z e^{\frac{V_c}{2K}(z-z')}\sinh\left[\frac{S_i}{2}(z-z')\right]q_i(z')dz'\nonumber\;;\\
S_i\!\!&\equiv&\!\!\sqrt{\frac{V_c^2}{K^2} \!+\! 4\frac{\zeta_i^2}{R^2}} ~~;~~A_i\!\!\equiv\!\! 2h \Gamma_{\rm tot} \!+\! V_c \!+\! KS_i\coth\left(\frac{S_iL}{2}\right).\nonumber 
\end{eqnarray}
In the above expression, $q_i(z)$ are the Fourier-Bessel 
coefficients of the source term ${\cal S}_{\rm prim}(r,z)$.
Some numerical issues about the convergence of the series, when using cuspy
dark matter halo, are underlined in App.~\ref{app:num_inst}.

    \subsection{Dark matter profile}
\label{subsubsec:Dark}
The primary exotic source term follows the generic form
\begin{equation}
 {\cal S}_{\rm prim}(r,z)= Q^{\bar{p}}(E) \times f_{\rm Source}(r,z)\;,
\end{equation}
where the spatial dependence of the source term is normalized to 1 at Solar position. For evaporation
of Primordial Black Holes or for SUSY-like particles, $f_{\rm Source}(r,z)$ 
is equal to
\[
  f_{\rm Dark}(r,z) \quad {\rm or} \quad f^2_{\rm Dark}(r,z)\;,
\]
where $f_{\rm Dark}(r,z)$ is the dark matter distribution profile, also 
normalized to 1 at Solar position. Any information such as, e.g., the relic density, the
fraction of the dark matter halo filled by the new candidate or
any quantity related to the dark matter candidate is
irrelevant at this stage: it is absorbed in the energy dependent term $Q(E)$.
The right factors should be implemented accordingly
when using the approximate formulae.

\begin{table}[!t]
\begin{center}
{\begin{tabular}{@{}|l|c|c|c|c|@{}}
\hline
Halo model & $\alpha$ & $\beta$ & $\gamma$ & $r_c$ [kpc] \\
\hline
\hline
Cored isothermal
& {\aaa} 2 {\aaa} & {\aaa} 2 {\aaa} & {\aaa} 0 {\aaa} & {\aaa} 4
{\aaa} \\
NFW~\citep{1997ApJ...490..493N,2004MNRAS.349.1039N}
&        1        &        3        &        1        &
25       \\
Moore~(Moore et al., Diemand et al.)
&        1      &        3        &        1.2      &
30       \\
\hline
\end{tabular}}
\end{center}
\caption{Parameters to be plugged into Eq.~(\ref{eq:profile}) to obtain
various modelings of the dark matter distribution profile in the Milky Way.
\label{tab:indices}}
\end{table}
The following generic form is taken for the profile:
\begin{equation}
f_{\rm Dark}(r,z) = \left( \frac{R_{\odot}}{\sqrt{r^2+z^2}} \right)^\gamma \,
\left( \frac{ r_c^\alpha  + R_{\odot}^{\alpha}}{r_c^\alpha + \sqrt{r^2+z^2}^{\alpha}}
\right)^{\left( \beta - \gamma \right) / \alpha} \;\; ,
\label{eq:profile}
\end{equation}
where parameters are given in Tab.~\ref{tab:indices}.
The distance to the galactic center is set to
$R_{\odot}=7.5$~kpc, a value that several methods seem to
converge to~\citep{2006ApJ...647.1093N}. This is at variance with the 
usually recommended $R_{\odot}=8.0$~kpc (e.g.~\citealt{PDBook}),
but has no impact on the derived results below. Indeed,
$R_\odot$ is just another parameter, and our approximate formulae
will perform well for any user-preferred value $\approx 8.0$~kpc.

\section{Approximate formulae}
\label{sec:alternative}
The previous formulae (see Sec.~\ref{subsec:2Dfull}) can be tricky to
implement in practice (numerical inversion, tertiary contribution,
convergence issues), 
especially when energy redistributions are included.
In this section, energy gains and losses are discarded
and a simplified formalism
is presented for the calculation of exotic primary \pbar\ fluxes.
Some of the formulae given hereafter are well suited
for extensive computation.

       \subsection{Propagator (no side boundaries: $R\rightarrow\infty$)}
\label{subsec:propagator2D}
In~\citet{taillet-2004-}, the expression for the propagator corresponding
to the same 2D diffusion equation (without energy redistributions)
was extracted in the limit\footnote{The propagator for finite $R$ was presented
and used in~\citet{2003A&A...404..949M} to obtain the most likely {\em origin} of exotic \pbar\ detected
on Earth. This propagator relied on Bessel expansion. For practical use, $R\gg 1$ was required, so
that it was equivalent to the form presented below. However, in the context of a quick and
simple formulae, it is largely outperformed (in term of convergence properties) by the one presented
in this paper.} $R\to \infty$. Hence, the problem is invariant under radial translations
and we choose the location of the observer as the origin ($\vec{r}=\vec{0}$).
The propagator ${\cal G}(\vec{r}-\vec{r'})$ is solution of
\[
 \displaystyle \left\{ -K\triangle +V_c \frac{\partial }{\partial z} +2hn\delta(z)v\sigma\right\} {\cal G} = \delta(\vec{r}-\vec{r'})
\]
and reads (see App.~\ref{app:check-propag})
\begin{eqnarray}
\displaystyle {\cal G}(r,z) &=& \frac{\exp^{-k_v z}}{2\pi K L} \times \\\nonumber
\displaystyle &&\sum_{n=0}^\infty c_n^{-1}K_0(r \sqrt{k_n^2+k_v^2}) \sin k_n L
\, \sin k_n (L-z)
\label{kn}
\end{eqnarray}
where $K_0$ is the modified Bessel function of the second kind,
$k_n$ is solution of \begin{equation}
 2k_n \cos k_n L = -k_d \sin k_n L\;,
\end{equation}
$c_n$ is defined as
\begin{equation}
 c_n = 1 - \frac{\sin k_n L \cos k_n L}{k_n L}
\end{equation}
and
\[
 k_v \equiv V_c/(2K)\;.
\]
The quantity $k_d$ depends on the total destruction rate $\Gamma_{\rm tot}$
and the constant convective wind $V_c$:
\begin{equation}
k_d \equiv 2h \;\Gamma_{\rm tot}/K + 2 k_v\;.
\end{equation}

The flux at solar location is given by
\[
 N_\odot= 2\int_0^{2\pi}\int_0^L \int_0^R {\cal G}(r,\theta,z) {\cal S}_{\rm prim}(r,\theta,z) rd\theta drdz\;.
\]
where ${\cal S}_{\rm prim}(r,\theta,z)$ is the dark matter source term as seen from 
the Solar System.
Hence, a single (numerical) integration is needed in the propagator approach: this represents
a great benefit compared to the full 2D description given in Sec.~\ref{sec:2D}.

   \subsection{The 1D model}
   \label{subsec:1Dmodel}
Another way to tackle the simplified problem (neglecting energy redistributions) is the
use of a 1D model.
We now consider an infinite disk whose density and source distribution
do not depend on $r$.
The constant wind diffusion transport equation for a generic source
term ${\cal Q}(z,E)=q(z) Q(E)$ without energy redistributions reads
\begin{equation}
\label{1D-exotic-thin-disk}
 -K \frac{d^2N}{dz^2} +V_c\frac{dN}{dz} + 2h\delta(z) \Gamma_{\rm tot} N
 = q(z) Q(E) \;.
\end{equation}

As we do not take into account energy losses, 
the energy $E$ is decoupled from the spatial dependence $z$
so that it is omitted. In 1D--models, only $z$-dependence
is allowed. For simplicity, it is further assumed that the source term $q$
is constant throughout the Galaxy.
We refer the reader to the companion paper for details
of the calculation. The analytical solution reads
\begin{eqnarray}
   \left\{
   \begin{array}{ll}
\label{eq:prim-Vc-general}
 \displaystyle      N(z)=\frac{qL}{V_c}
\left\{
 \frac{(1+\alpha+\xi) \cdot(1-e^{-\alpha(1-\frac{z}{L})})}{\alpha+\xi(1-e^{-\alpha})}
 +\frac{z}{L}-1\right\}\vspace{4mm}
 \\
\displaystyle N(0)=\frac{qL}{V_c}\left\{
\frac{1-(1+\alpha)e^{-\alpha}}{\alpha+\xi(1-e^{-\alpha})}
\right\}
   \end{array}
   \right.
\end{eqnarray}
where $\xi=h\Gamma_{\rm tot} L/K$ and $\alpha=V_cL/K$.

\section{Results}
\label{sec:results}
In this section, the validity of the previous approximate formulae is checked. We first study the 
influence of energy redistributions within the 2D numerical framework and then compare the two simplified 
formulae to the 2D numerical model (without energy redistributions).

\subsection{Influence of energy redistributions in the 2D model}
\label{subsec:energy_Redistrib}
\begin{figure}[!t]
\begin{center}
\includegraphics[width=\columnwidth,angle=0]{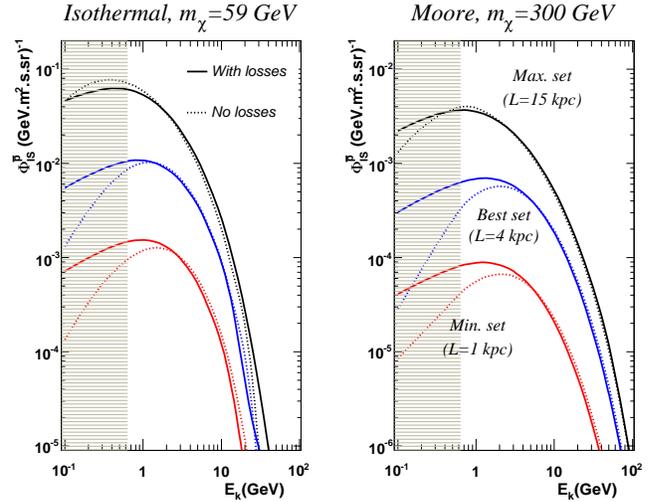}
\caption{From top to bottom,
the {\em max}, {\em best} and the {\em min} propagation sets (see Tab.~\ref{tab:param}).
The dashed lines correspond to the propagated fluxes without energy redistributions,
while solid lines correspond to the same propagated exotic fluxes, but with
energy redistributions (energy losses, reacceleration, tertiary source term)
taken into account. The shaded regions correspond to \pbar\ which cannot reach Earth (because of
solar modulation, see text). The two panels correspond to different dark matter profiles
and neutralino masses. 
\label{fig:losses_vs_nolosses}}
\end{center}
\end{figure}
The realistic dark matter profile (as described in Sec.~\ref{subsubsec:Dark})
is implemented in the full 2D--model (see Sec.~\ref{subsec:2Dfull}),
with and without energy redistributions. When energy redistributions are neglected,
the quantity ${\cal N}_i(0)$ to plug in Eq.~(\ref{Ni_nocal}) becomes
the quantity ${\cal N}^\star_i(0)$ defined in Eq.~(\ref{Ni_cal}).
We use the generic term \error~for
the difference between solutions calculated with and without energy
redistributions. 

\subsubsection{Expected effect of energy redistributions}
As energy redistributions act on the spectrum, it is not obvious to
disentangle which of the following input parameters---source spectrum, profile, transport
coefficients---is the most crucial regarding \error. On a general footing,
energy losses and tertiary contributions tend to replenish the low energy tail in case
of a dropping flux at these same energies (this has a particularly important
impact on the standard secondary \pbar, see \citealt{2001ApJ...563..172D}).
On the other hand, reacceleration tends to
smear the spectrum at low and high energy in case of a flux peaking at GeV energy.
Taking all these points into account, the following remarks---regarding energy
redistributions---can be made:
\begin{description}
 \item[Source spectrum:] in most cases, PBH-like source spectra do not decrease
       at low energy (e.g., Fig.~3 of~\citealt{2002A&A...388..676B}), whereas SUSY-like candidates
       are quite flat (e.g., Fig.~6 of~\citealt{2004PhRvD..69f3501D}). Hence, taking various
			 source spectra is not expected to play a major role in \error.
 \item[Dark matter profile:] changing
       the spatial distribution does not affect the spectral distribution: it
			 only changes the normalization of the propagated spectrum.
       Hence, no significant effect on \error~is expected when switching
			 between SUSY-like to PBH-like source terms.
 \item[Propagation parameters:] Large winds dominate over diffusion as the energy
			 diminishes: they decrease low energy fluxes, making them more sensitive to energy
			 redistribution effects. These large wind happen to be associated with a small halo size
			 and a small reacceleration parameter. This effect is larger than the two others (see below).
\end{description}

\subsubsection{Behavior of \error}
Figure~\ref{fig:losses_vs_nolosses} displays the two resulting
fluxes for various input configurations. 
The decrease of the flux observed at low energy for the best and minimal
propagation sets is due to the Galactic
wind. For the smallest $L$, the propagated flux is steeper
(because of larger wind values, see Tab.~\ref{tab:param}), hence more
affected by energy redistributions.  Notice that only for $L=15$~kpc
(corresponding to $V_c=5$~km~s$^{-1}$),
can the shape of the source spectrum be truly appreciated.
As expected, the \error~calculated when  varying the neutralino mass or
the dark matter profile is less important.

The \error~can be as large as a factor of $\sim 5$ at very low energy, but
we remind that these are \pbar\ at IS energies, which cannot reach
Earth because of solar modulation (shaded region in Fig.~\ref{fig:losses_vs_nolosses}).
Indeed, the top of atmosphere energy $E_{\rm TOA}$ is related to
the interstellar energy $E_{\rm IS}$ by $E_{\rm IS}= E_{\rm TOA}+\phi$
where $\phi$ is the modulation parameter $\in[500-1000]$~MV \citep{1987A&A...184..119P}.

Above a few hundreds of MeV (lowest IS energy reachable for low solar
activity), the difference obtained is always $\lesssim 50\%$. Above a few tens
of GeV, energy redistributions are always negligible, so that \error~tends to
zero. In between, the only situation when \error~can be large corresponds
to a steep spectrum dropping rapidly to zero, which is uninteresting
in the perspective of detecting any excess.

\subsubsection{Summary}
The model neglecting energy redistributions is accurate
enough to determine exotic fluxes. 
This result is in agreement with the intuitive idea that exotic species,
in order to reach us, cross the gaseous disk less often that standard ones,
implying that they are less subject to energy redistributions.

\subsection{Approximate formulae vs 2D reference model}
\label{subsec:Comparison}

\subsubsection{Propagator}
As expected, the propagator (see Sec.~\ref{subsec:propagator2D})
and the Bessel expansion (see Sec.~\ref{subsec:2Dfull}) formalisms
are consistent, except when the
halo size has the same extension of the radial extension of the Galaxy $R$.
In that case, we no longer have $L\ll R$ and the side boundary
starts affecting the flux, as discussed
in~\citep{2003A&A...402..971T,2003A&A...404..949M}.

       \subsubsection{1D--model}
\begin{figure}[!t]
\begin{center}
\includegraphics[width=0.49\columnwidth,angle=0]{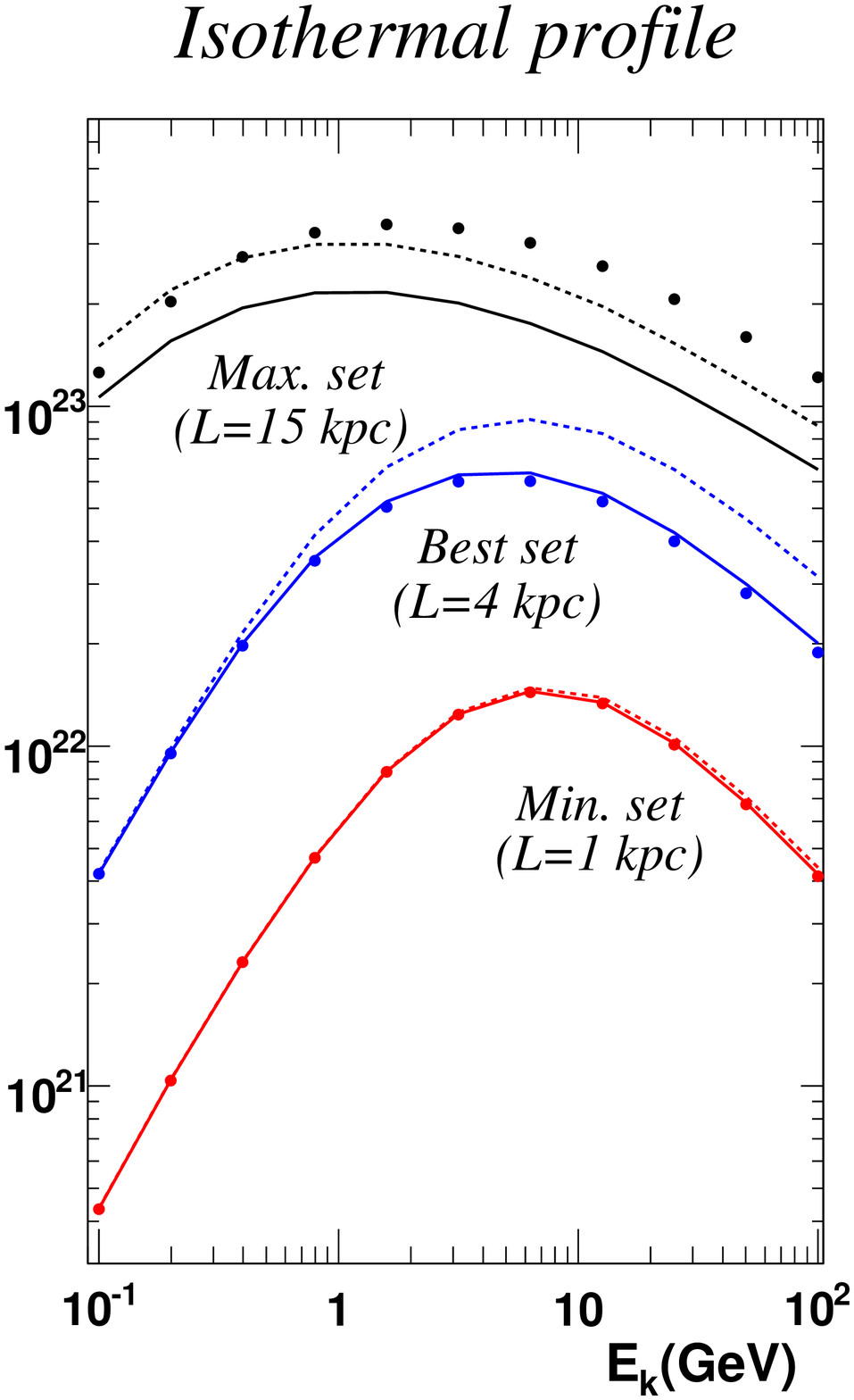}
\includegraphics[width=0.49\columnwidth,angle=0]{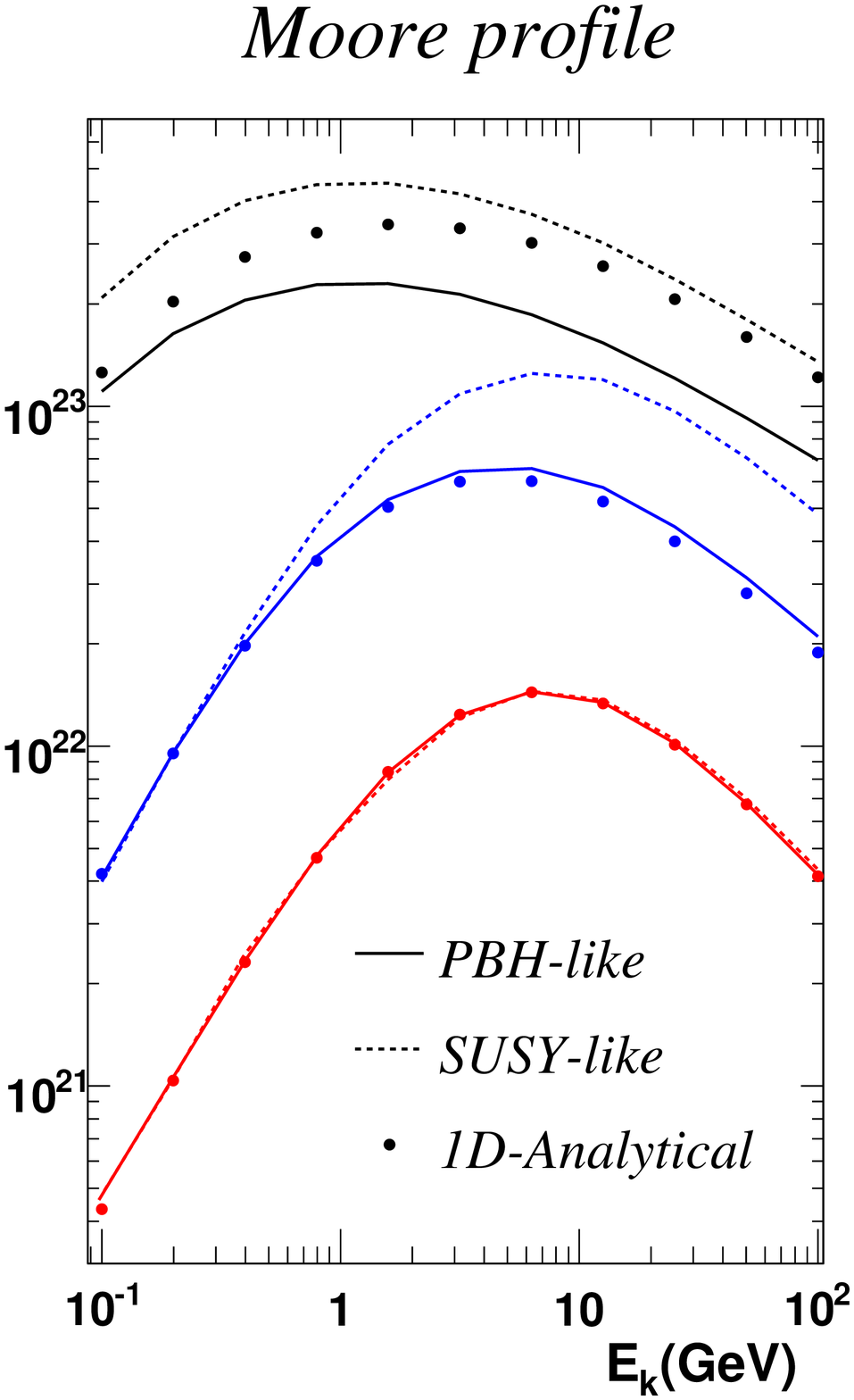}
\caption{Both figures correspond to the ratio of the \pbar\ exotic flux to its source term (in arbitrary units)
as a function of the kinetic energy. Lines: 2D--semi-analytical model, no energy redistributions
(Sec.~\ref{subsec:2Dfull}) for PBH-like (solid lines) or SUSY-like (dashed lines) sources.
Symbols: Eq.~(\ref{eq:prim-Vc-general}), 1D--model (independent of the dark matter profile). The three sets of propagation parameters
are those of Tab.~\ref{tab:param}. Left panel: Isothermal profile; right panel: Moore profile.
\label{fig:2D_vs_1D_ISO_MOORE}}
\end{center}
\end{figure}
Figure~\ref{fig:2D_vs_1D_ISO_MOORE} compares
the 1D--model (at $z=0$) to the 2D semi-analytical model:
\begin{itemize}
       \item For PBH-like sources, as long as $L\lesssim 7$~kpc, the 1D--model provides a 
			 good description of exotic \pbar\ fluxes.
       \item For SUSY-like sources, 1D--models are valid for very small halo size $L$ only.
\end{itemize}
This good agreement is easily understood if we remember that the closest boundary
defines a cut-off distance: anti-nuclei from sources located further away are
exponentially suppressed~\citep{2003A&A...402..971T,2003A&A...404..949M}. Hence,
we are only sensitive to the average value of $f_{\rm Dark}$ within
this cut-off distance, close to one at solar position, as taken in the 1D--model. 
As soon as contributions close to the Galactic center are less suppressed,
this equivalence breaks down; all the more for SUSY-like source terms compared to
PBH-like source terms (because of the square in $f_{\rm Dark}$).

\section{Numerical routines}
\label{sec:summary}

Taking advantage of the two simplified formalisms depicted above (propagator
and 1D model),
routines are provided\footnote{http://wwwlapp.in2p3.fr/$\sim$taillet/mtc/mtc\_code.tar}, to evaluate the \pbar\ flux
from any input dark matter profile, source spectrum,
and propagation parameters.
The main routine returns
the ratio of the propagated IS exotic flux (for a user-specified source term) to the
standard \pbar\ IS flux, i.e.
\[
 \mho \equiv \Phi^{\bar{p}~{\rm - IS}}_{\rm exotic}/\Phi^{\bar{p}~{\rm - IS}}_{\rm standard}\;.
 \]
The input parameters are the following:
\begin{itemize}
       \item Propagation parameters: $K_0$ (kpc$^2$~Myr$^{-1}$), $\delta$, $V_c$ (km~s$^{-1}$) and $L$ (kpc);
       \item Exotic source spectrum: $Q^{\bar{p}}_{\rm exotic}(E_k)$ (GeV$^{-1}$~m$^{-3}$~s$^{-1}$)
       and the corresponding kinetic energy $E_k$ (GeV);
       \item Dark matter profile: the parameters $\alpha$, $\beta$ and $\gamma$ (see~\ref{tab:indices}) as
       well as the source type (SUSY-like or PBH-like).
\end{itemize}
The resulting fluxes, in (GeV~m$^{2}$~s~sr)$^{-1}$, are obtained at solar position $(R_\odot,0)$.
The IS background \pbar\ spectrum is described
using a fit function (see Sec.~\ref{app:fit}).
We chose to express all fluxes as IS quantities to free the user
from using a modulation routine. There are uncertainties
associated with the choice of the modulation/demodulation scheme,
but these are not larger than the error made using approximate
formulae (see \citealt{2002ApJ...565..280M} for a discussion
on solar modulation).

We remind that the range of propagation parameters given in
  Tab.~\ref{tab:param} implies that large
uncertainties remain on exotic fluxes (without mentioning the fact that
the constant wind model is probably not the {\em definitive}
model for the Galaxy). To make conservative estimates,
we recommend the reader to use the $min$ set of propagation parameters
associated with the condition $\mho\gtrsim 1$ to exclude an exotic
model. Regarding the most likely \pbar\ exotic flux in this model, it is given by
the $best$ set ($L=4$~kpc, see Tab.~\ref{tab:param}). In any case,
it has to be kept in mind that energy smaller than $\sim 600$~MeV IS correspond to
\pbar\ which are almost never detected on Earth.

\section{Conclusion}
\label{sec:theEnd}
Given the present accuracy of propagation parameters,
exotic anti-proton fluxes suffer large uncertainties (see also the companion paper).
This limits the benefit of a detailed calculation, which
can be expensive in term of computational power, when
repeated thousands of times.

We showed that discarding energy redistributions (coulomb, ionization, adiabatic losses, reacceleration
and tertiary contribution) provides exotic \pbar\ fluxes accurate at the level of $\lesssim 50\%$
for IS energies greater than $\sim 1$~GeV (leading to $\gtrsim 100$ MeV energies once modulated).
This opens the possibility to use one of the following simpler formulae:
\begin{itemize}
       \item the approximate 1D--model, which gives the correct
         flux as long as $L\lesssim 7$~kpc
       for PBH-like sources, or $L\lesssim 1$~kpc for SUSY-like sources;
       \item the 2D-propagator.
\end{itemize}
The advantage of the 1D--model is its compactness and simplicity
(no dependence on the dark matter profile),
while the second approach is more accurate.
The former could be preferred in situations where speed is a decisive factor,
e.g. when scanning a large parameter space as in {\sc susy} studies.
Moreover, in order to be conservative in exclusion studies, small
halo size should be taken. In particular conservative estimates
are made when using the 3$^{\rm rd}$ line of Tab.~\ref{tab:param}:
this is exactly the regime where 1D--formulae hold. We also supply numerical
routines that could be easily implemented in indirect detection codes.

So far, these formulae are valid only in the framework
of the constant wind model. Although all the discussions
were developed with respect to exotic \pbar\ fluxes, all the results can
be transposed almost without any modification to the \dbar\ case
(the total cross section has to be taken accordingly to the
species considered).


\section*{Acknowledgments}
We thank E. Nezri for useful discussions held during the GDR-SUSY meeting,
and J. Lavalle and F. Donato for a careful reading of the manuscript.

\appendix

\section{Check of propagator formulae}
\label{app:check-propag}
The validity of the propagator is checked in a simple case,
by integrating it over a constant source term $q(r,z)=1$ (1D formula).

  \subsection{Purely diffusive regime: $V_c=0$}

\subsubsection{No spallations, $k_d = 0$} In that case, $k_n \equiv (n+1/2) \pi$.
Integration over $z$ (using $q(r,z)=1$) leads to
\begin{equation}
\int_{-L}^L dz \, {\cal G}(r,z) = \frac{1}{\pi K L} \sum_{n=0}^\infty
\frac{(-1)^n}{k_n} K_0(k_n r)\;,
\end{equation}
and integration over the disk $\int_0^R 2\pi r \, dr$ gives
$$N(0) = \frac{2}{K L} \sum_{n=0}^\infty \frac{(-1)^n}{k_n^3}
\int_0^{k_nR} y \, K_0(y)\, dy\;.$$
Using the properties $\int y \, K_0(y)\, dy = -y K_1(y)$
and $y K_1(y) \sim 1$ when $y \rightarrow 0$,
$$N(0) = \frac{2}{K L} \sum_{n=0}^\infty \frac{(-1)^n}{k_n^3}
\left[1 - k_n R \, K_1(k_n R)\right]\;.$$
Finally, for $R\gg 1$, the expression becomes
$$N(0) = \frac{2}{K L} \sum_{n=0}^\infty \frac{(-1)^n}{k_n^3}
= \frac{2L^2}{\pi^3 K}\sum_{n=0}^\infty \frac{(-1)^n}{(n+1/2)^3}\;.$$
The infinite sum has the value $\sum (-1)^n/(n+1/2)^3 = \pi^3/4 \approx 7.7515$, so that
$$N(0) \;\;= \;\;\frac{L^2}{2 K} \quad \equiv  N^{\rm 1D-model}(0).$$

Note that, in the infinite sum, the first term $n=0$ is equal to 8, so that
it only slightly overshoots the exact value.
Hence, only a few terms are required in the sum to converge quickly
to the correct value.

\subsubsection{With spallations, $k_d=2h\Gamma_{\rm tot}/K$}
There is no longer a simple expression for $k_n$. Following a similar derivation
as the one above, we find
\begin{eqnarray}
\label{eq:propagator_no_Vc}
N(0) = \frac{4}{K k_dL}
\sum_{n=0}^\infty \left(1 - \frac{\sin k_n L \cos k_n L}{k_nL}\right)^{-1}\nonumber\\
\times \frac{\cos k_n L}{k_n^2} \left[ \cos k_nL -1 \right]\;.
\end{eqnarray}

A numerical check of this sum confirms that Eq.~(\ref{eq:propagator_no_Vc})
equals the 1D--model, i.e. Eq.~(\ref{eq:prim-Vc-general}) where $V_c=0$. This means that
\[
\sum_{n=0}^\infty
\frac{   \cos (k_n L) \left[ \cos (k_nL) -1 \right]}{ k_nL\left(k_nL - \sin (k_n L) \cos (k_n L)\right)}
= \frac{k_d L}{4(2+k_dL)}\;.
\]

  \subsection{Diffusive/convective regime: $V_c\neq 0$}
The integration leads to quite similar results as for the previous
case (with spallations). A numerical check confirms that
Eq.~(\ref{eq:prim-Vc-general}) is recovered.

Note that the integration should be performed for sources
located at all $r$, with $r\rightarrow\infty$. In practice,
it is sufficient (and it saves a lot of computational time)
to integrate only from 0 to $10L$ (or from 0 to $\min(10L, 10 L_{\rm eff})$)
in the case of a convective wind, where $L_{\rm eff}\equiv K/V_c$.
In that case, a grid
of $\sim 25$ points for all coordinates ($r$, $z$ and $\theta$) is sufficient
to reach the correct result.


\section{Numerical instabilities in the 2D model}
\label{app:num_inst}
Dealing with Bessel functions is a source of numerical instabilities.
These occur if i) $R\gg L$ or ii) too many Bessel orders are used,
or even, in the case of cuspy dark matter halo, if too few orders are used. As the propagator
formulation allows cross-checks, we take the opportunity to give a few recommendations
regarding the parameters to use:
       \begin{enumerate}
               \item always keep $R$ in the range $5L^\star \lesssim R\lesssim 10L^\star$, with 
							 $L^\star= \text{min} (L,K/V_c)$.
							 The limit $10L^\star$ is set
               to avoid numerical instabilities
               while the limit $5L^\star$ allows to avoid boundary effects~\citep{2003A&A...404..949M}
               (the boundary in $R$ decreases the flux at most to a few tens of percent for a Moore profile).
               \item For a smooth profile (e.g. Isothermal), $20$ Bessel functions give an excellent convergence.
               \item For NFW and Moore profiles in the case of SUSY-like candidate ($50$ Bessel functions are sufficient for
               PBH-like candidate), following a procedure inspired by~\citet{2005PhRvD..72f3507B}, we replace in the
               calculation the standard profile, by
               \begin{eqnarray}
    g^2_{\rm Dark}(r)\!=\!\left\{
   \begin{array}{ll}
                       \displaystyle \!\!r_{\rm th}^{2\gamma} \cdot f^2_{\rm Dark}(r_{\rm th}) \cdot
                               \pi^2\Upsilon \cdot \!\frac{\sin(\pi r/r_{\rm th})}{\pi r/r_{\rm th}} \quad \!{\rm if~} r\leq r_{\rm th} \;;\\
                       \displaystyle \!f^2_{\rm Dark}(r) \quad {\rm otherwise};
   \end{array}
   \right.
       \end{eqnarray}
       with $\Upsilon=(3-2\gamma)^{-1}$ and $\gamma$ is the slope from Tab.~\ref{tab:indices}.
       In these expressions, $r\equiv \sqrt{r_{\rm cyl}^2+z^2}$ denotes
       the spherical coordinate. In any case, we find that, setting $r_{\rm th}=400$~pc, $50$ Bessel functions
       give a good accuracy, while 100 functions allow to reach an excellent convergence.
       \end{enumerate}

\bibliography{mtc_split1}
\end{document}